\documentclass{sig-alternate-05-2015}

\usepackage{flushend}
\usepackage{subcaption}
\usepackage{color}
\usepackage[utf8]{inputenc}
\usepackage{multirow}
\usepackage{newfloat}

\DeclareFloatingEnvironment[fileext=lop]{query}

\newcommand{\ttt}[1]{\texttt{#1}}
\title{Query Expansion via structural motifs in  Wikipedia Graph}
\numberofauthors{3} 
\author{ \alignauthor Joan Guisado-Gámez\\ 
	\affaddr{DAMA-UPC}\\ 
	\affaddr{Universitat Polit\`ecnica de Catalunya}\\
	\email{joan@ac.upc.edu} 
	\alignauthor Arnau Prat-Pérez\\ 
	\affaddr{Sparsity-Technologies}\\ \affaddr{Barcelona} 
	\email{arnau@sparsity-technologies.com}    \alignauthor Josep Lluis Larriba-Pey\\ 
	\affaddr{DAMA-UPC}\\ \affaddr{Universitat Polit\`ecnica de Catalunya}\\
	\email{larri@ac.upc.edu} }   

	\newfont{\mycrnotice}{ptmr8t at 7pt}
	\newfont{\myconfname}{ptmri8t at 7pt}
	\let\crnotice\mycrnotice%
	\let\confname\myconfname%

\begin{document}
	\doi{}
	\conferenceinfo{PRE PRINT}{}
	\isbn{}

	\maketitle

	\begin{abstract}
		The search for relevant information can be very frustrating for users 
		who, unintentionally, use too general or inappropriate keywords to 
		express their requests.  To overcome this situation, query expansion 
		techniques aim at transforming the user request by adding new terms, 
		referred as expansion features, that better describe the real intent of 
		the users.  We propose a method that relies exclusively on relevant 
		structures (as opposed to the use of semantics) found in knowledge bases 
		(KBs) to extract the expansion features.  We call our method Structural 
		Query Expansion (SQE). The structural analysis of KBs takes us to 
		propose a set of structural motifs that connect their  strongly related 
		entries, which can be used to extract expansion features.  In this paper 
		we use Wikipedia as our KB, which is probably one of the largest sources 
		of information. SQE is capable of achieving more than 150\% improvement 
		over non expanded queries and is able to  identify the expansion 
		features in less than 0.2 seconds in the worst case scenario. Most 
		significantly, we believe that we are contributing to open new research 
		directions in query expansion, proposing a method that is orthogonal to 
		many current systems.  For example, SQE improves
		pseudo-relevance feedback techniques up to 13\%.
	\end{abstract}

	\section{Introduction} 

	In a typical information retrieval scenario, users express their needs of 
	information with a set of keywords. However, \textit{vocabulary mismatch} 
	between the keywords and the documents to be retrieved entails poor results 
	that do not satisfy the user needs~\cite{metzler2007similarity}. Poor 
	results also arise from the \textit{topic inexperience} of the users because  
	they are not  often familiar with the vocabulary and use too general 
	keywords.


	Query expansion techniques aim at improving the results achieved by the user
	request by means of introducing new terms, commonly called \textbf{expansion
	features}. Thus, the challenge is to select those expansion features that
	are capable of improving the results the most.  A bad choice of
	expansion features may be counterproductive.  There are different families
	of expansion techniques, which differ in the way they
	acquire the expansion features. One such family consists in using knowledge 
	bases (KBs). To 
	the best of our knowledge, those rely on some kind of text analysis, such 
	as explicit semantic analysis~\cite{aggarwal2012}, or are based on other 
	existing query expansion techniques such as pseudo-relevance 
	feedback~\cite{Arguello2008}. However, as we show here, the underlying network structure of 
	KBs has been barely exploited. In~\cite{guisado2013Icmr}, we 
	presented a proof of concept in which we used the network structure of a 
	KB. Although we achieved good results, the algorithms that we used were 
	based on~\cite{PerezDBL12}, which is a metric designed for social network 
	analysis and hence, it does not exploit the particular KBs structures.


	In this piece of work, we propose Structural Query Expansion (SQE), a new query expansion strategy to 
	exploit KBs relying exclusively on the structural relationships among  data.  
	For that purpose we present a structural analysis of a KB, Wikipedia, and 
	from this analysis we define a set of structural motifs that are capable of 
	identifying reliable expansion features. In other words, we show that it is 
	not necessary to use semantic analysis, but just to look at the structure of 
	KBs to obtain good expansion features.

	\subsection{General Overview}

	The goal of this paper is to improve a typical query expansion process.  
	First, the system receives a user request, which is the \ttt{input} of the 
	system. Then, it identifies a set of expansion features that better 
	describe the user intent.  Finally, it builds an expanded query that is used 
	by the search engine to  retrieve the results.

	In order to identify the expansion features, we use a KB.  KBs consist of a 
	set of entries, each of which describes a single concept, and that relate to 
	each other forming a graph, where the nodes represent the entries and the 
	links their relationships. In this paper we propose SQE,  a novel approach 
	that relies exclusively on exploiting the structure of the KBs graph with no 
	need for analyzing their content in any way. 

\begin{figure}[]
		\centering
		\includegraphics[width=\linewidth]{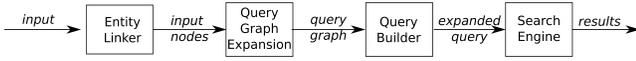}
		\caption{Search process with Wikipedia query expansion.}
		\label{fig:pipeline}
	\end{figure}

	The proposed approach follows the pipeline depicted in 
	Figure~\ref{fig:pipeline}. The \ttt{Entity Linker} receives the input, 
	usually expressed as a set of keywords. Then, it identifies the 
	\textbf{entities}, which are the real world concepts that appear in the 
	input. Finally, it matches these entities with nodes of the KB graph. We 
	refer to these nodes as the \texttt{input nodes}.  

	The \ttt{Query Graph Expansion} is in charge of selecting those nodes of the 
	KB graph that are tightly linked with the input nodes. From a 
	structural analysis of the KB graph, we describe a set of 
	structural motifs that allow deciding which nodes to select.  Out of those, 
	we induce a subgraph, which we name the \ttt{Query Graph}, that is a graph 
	representation of the original user request.  This is the main contribution 
	of this paper and it is described in detail in 
	Section~\ref{sec:StructuralQueryExpansion}.

	The query graph is used by the \ttt{Query Builder} to extract the \textbf{expansion 
	features} out of its nodes. Also, it builds the expanded 
	query with the input, the entities and the expansion features.  Then, the 
	expanded query is issued to the \ttt{Search Engine} that uses it to retrieve 
	the results. 


  In the particular case of this paper, we use Wikipedia as our KB. Wikipedia
  has two types of entries: \textbf{Article} and \textbf{Category}.  Articles
  are used to link the entities with the input nodes because each 
  describes a single topic.  Articles and categories form a network in which
  articles can link to other articles and must belong to, at least, one
  category. In addition, each category can belong to one or more general
  categories. From the analysis of this network, we propose the structural
  motifs to create the query graphs. Finally, the expansion features are extracted out
  of the query graph articles, in particular from their title,  that according
  to the Wikipedia's edition rules, must be recognizable, natural, precise,
  concise and consistent.  

	According to the conducted experiments, SQE is able to 
	obtain statistically significant improvements of more than 150\% over the
	non-expanded queries. 
	Moreover, SQE does not incur in a 
	relevant overhead, by running in the order of a few tenths of a second at 
	most.

	SQE is orthogonal to many other existing techniques. For example, in 
	Section~\ref{sec:experiments} we show that combining SQE with 
	pseudo-relevance feedback achieves 13.68\% improvement in the quality of the 
  results.  We believe that we are contributing to open new research directions 
that  consist in understanding and exploiting the particular structures 
	that characterize KBs in order to extract relevant expansion features, which 
	can be potentially combined with other strategies.  
	
	\subsection{Contributions}
	The contributions of this paper can be summarized as follows:

	\begin{itemize}
    \item We open new research directions consisting in exploiting the 
			structure of KBs to identify structural motifs that allow  
			connecting semantically related entries.  

		\item We use Wikipedia as our KB, we analyze its structure 
			and identify its relevant structural motifs.  
		
		\item We propose SQE, which is orthogonal to existing strategies and it 
			is executed in sub-second times, which are not a burden for the search process.

		\item We test the capabilities of SQE with three different 
			datasets: Image CLEF 2011, CHiC CLEF 2012 and CHiC CLEF 2013 and 
			validate the results with statistical significance analysis.
	\end{itemize}

	\vspace{20mm}
	{\begin{center}  \textbf{Paper Organization}\end{center}} The rest of 
  the paper is organized as follows. 
	In Section~\ref{sec:StructuralQueryExpansion}, we propose SQE from the analysis of the particular structures of Wikipedia.
	In Section~\ref{sec:experimentalSetup}, we give details about the Entity Linker, the Query Builder and the Search Engine as well as other details to understand the experiments. 
	In Section~\ref{sec:experiments}, we show the results achieved by SQE. 
	In Section~\ref{sec:relatedWork} we summarize the state of the art regarding query expansion, and we contextualize our work in it. 
	Finally, in Section~\ref{sec:conclusions}, we conclude and outline our future work.

	\section{Structural Query Expansion}\label{sec:StructuralQueryExpansion}
	As a starting point for the work in this paper we take~\cite{guisado2015}, 
	where we use an information retrieval dataset, Image CLEF, to create a 
	ground truth.  For that purpose, for each Image CLEF request we extract the 
	entities of its valid documents and match them with the articles in 
	Wikipedia. From this, we induce the ground truth query graphs.  Then, we 
	analyze the query graphs in order to identify the relevant structures that 
	relate the original input nodes to the set of articles that are good to 
	extract reliable expansion features. In the rest of this section, we 
	elaborate on these concepts to motivate the proposal of SQE based on query 
	graphs.

	\subsection{Wikipedia Structure Analysis}

	From the analysis of the query graphs, we have learned that, in general, they 
	are disconnected graphs composed by a moderately large connected 
	component.  Also,  that the largest connected component, and thus the 
	one that provides more expansion features, contains, in general, all the 
	input nodes for the ground truth query graph. 
	This supports the strategy of using the Entity Linker to identify the 
	input nodes, and then build the query graphs
	and also that Wikipedia encodes, in its 
	structure, the information to select reliable expansion features. 
	Finally, we have observed that the number of expansion features introduced is 
	very variable, it goes from 0 to 176.  This result means that we cannot 
	establish a unique and good number of expansion features as it 
	depends on the particular nature of each query.


	\begin{figure*}[]
		\centering
		\begin{centering}
			\begin{subfigure}[b]{.3\linewidth}
				\includegraphics[width=1\linewidth]{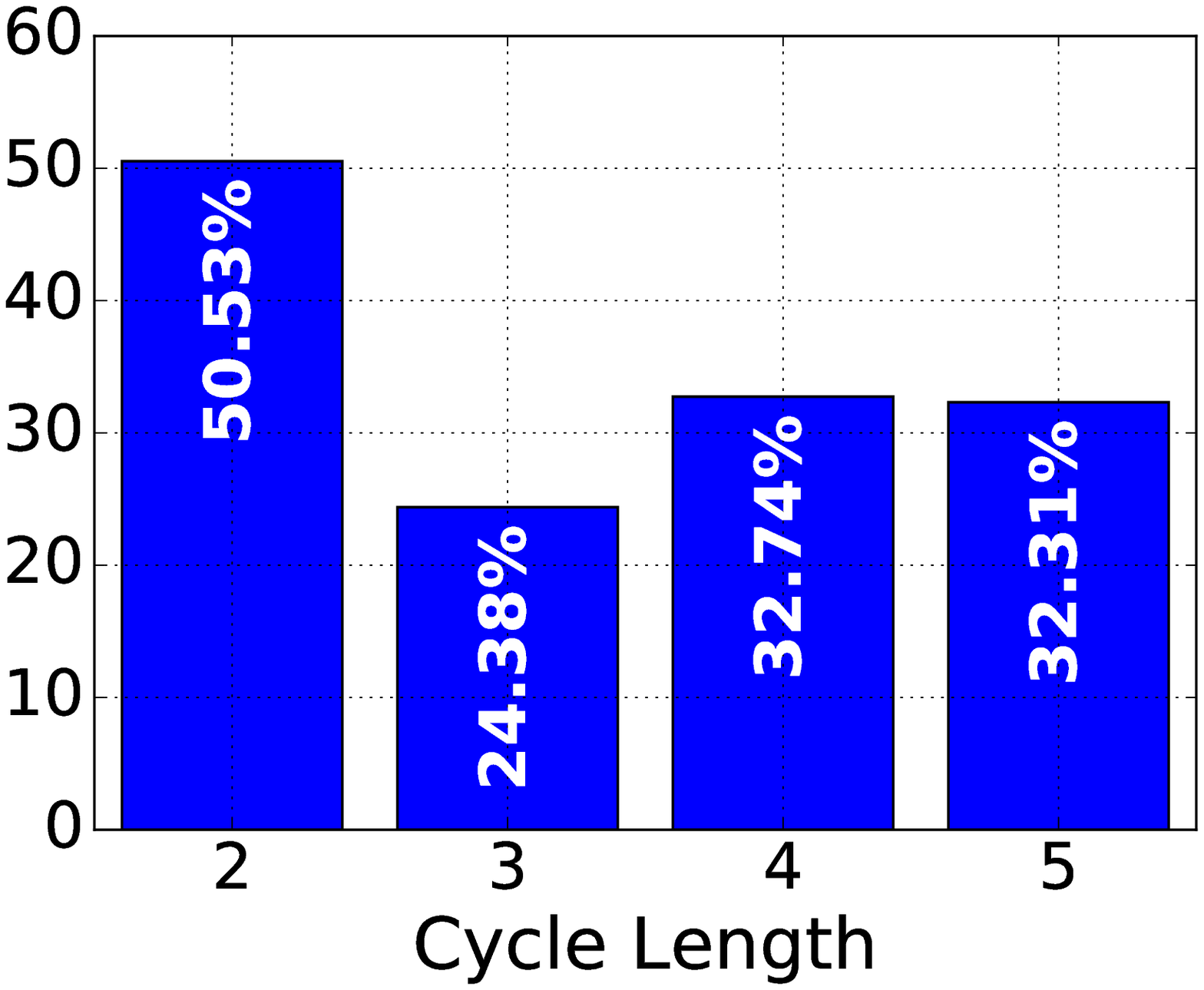}
				\caption{Contribution of cycles.}
				\label{fig:CycleLenghtContribution}
			\end{subfigure}
			\begin{subfigure}[b]{.3\linewidth}
				\centering
				\includegraphics[width=1\linewidth]{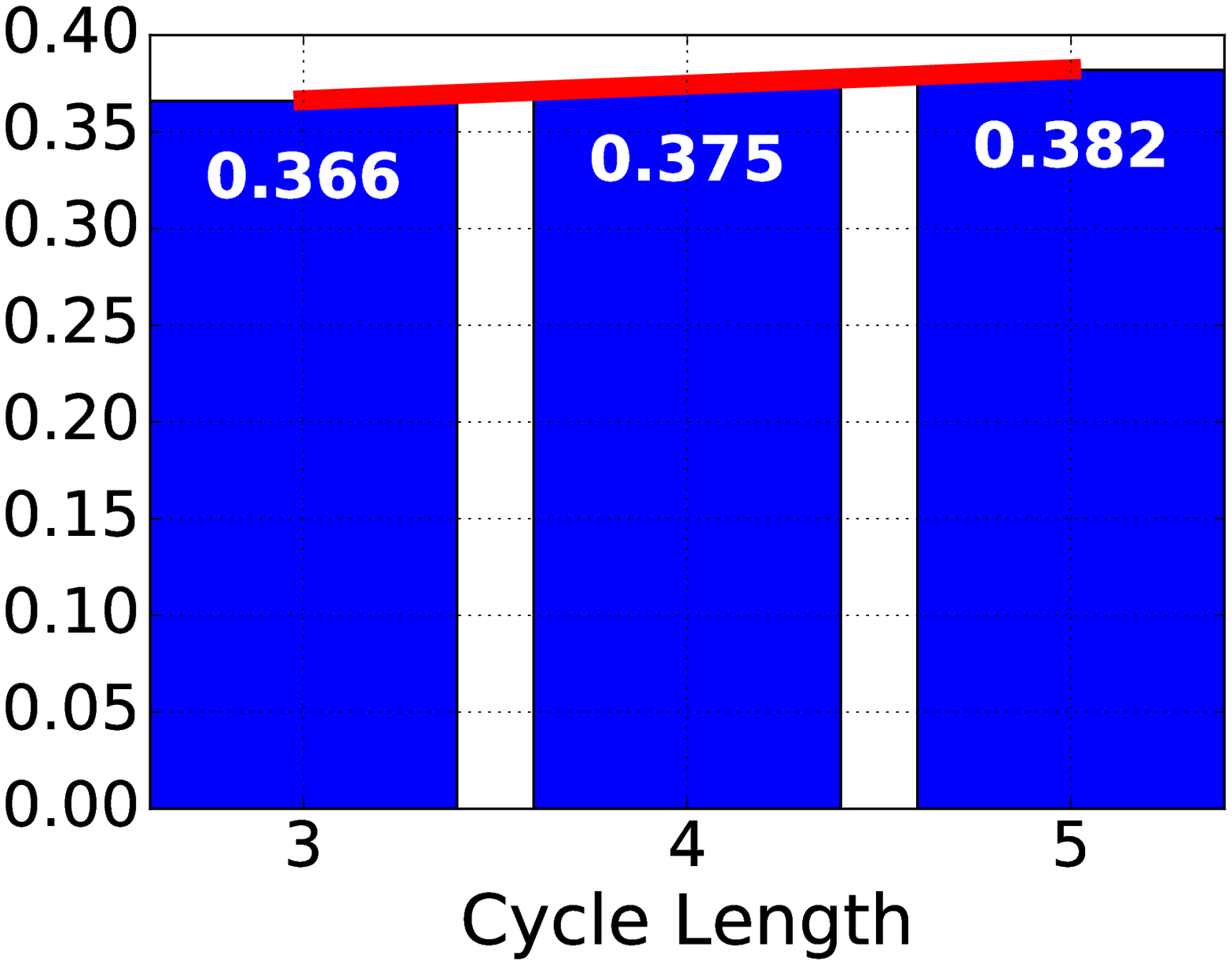}
				\caption{Average category ratio.}
				\label{subfig:categoryRatio}
			\end{subfigure}
			\begin{subfigure}[b]{.3\linewidth}
				\centering
				\includegraphics[width=1\linewidth]{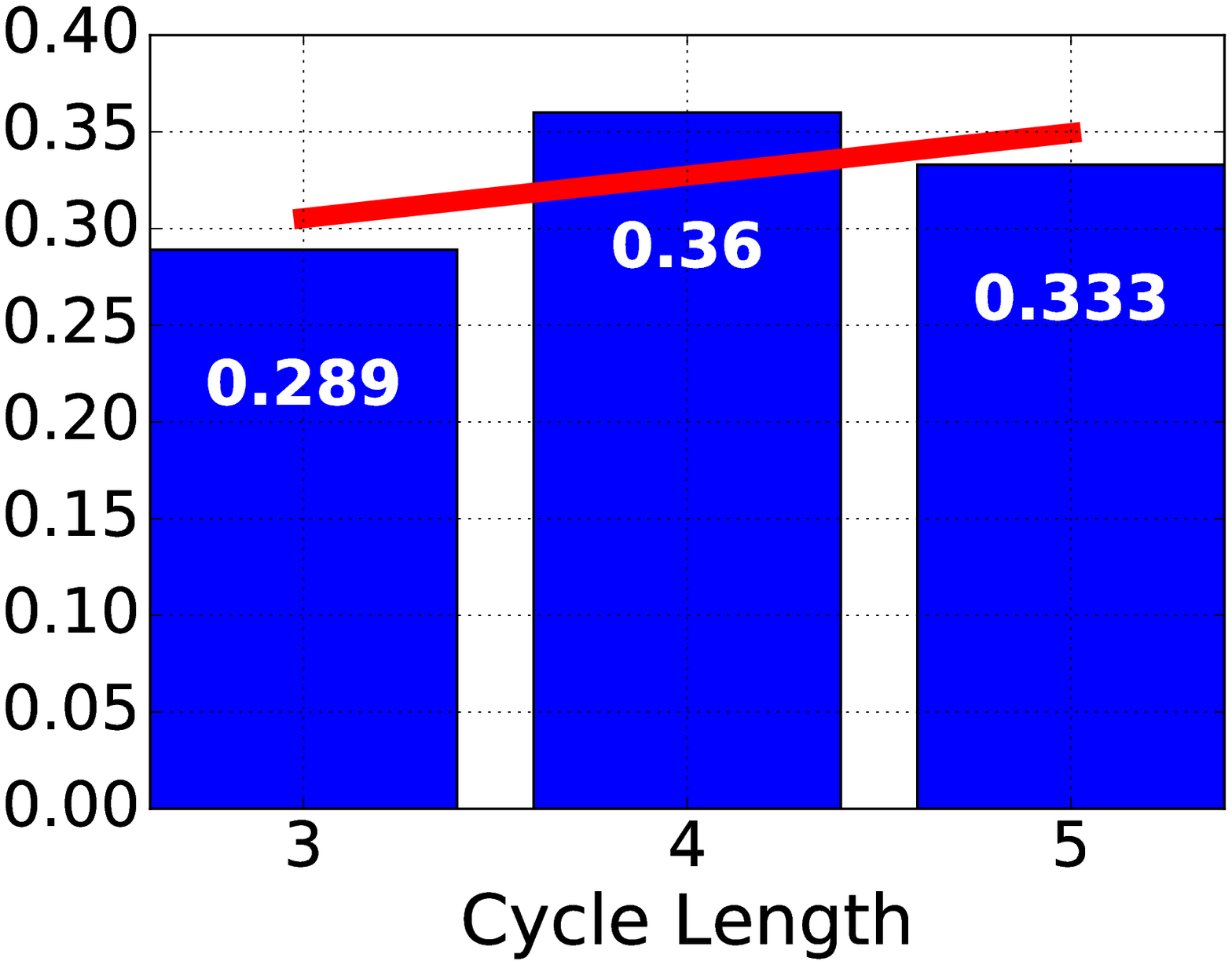}
				\caption{Average density of extra edges.}
				\label{subfig:densityOfExtraEdges}
			\end{subfigure}
			\caption{Characteristics of cycles of length 2, 3, 4 and 5.}
			\label{fig:CyclesCharacteristics}
		\end{centering}
	\end{figure*}
	
   	When we turn to analyze the relationship between the input nodes and the  articles in the query 
	graphs, we observe that cycles are very relevant to explain the 
	structure.
	In this case, cycles are defined as a closed 
	sequence of nodes, either articles or categories, with at least one edge 
	among each pair of consecutive nodes. The goal of our research is to 
	identify a set of characteristics of the cycles (length, ratio of articles 
	and categories and density of edges) that allow us to identify the 
	articles in those query graphs. 

	Concerning the length of the cycles, we evaluate the precision achieved by 
	using the expansion features extracted out of the articles reachable through 
	cycles of length, 2, 3, 4 and 5 isolated from the rest of the nodes of query 
	graphs. In Table~\ref{table:cycleResults}, we summarize those results. We 
	realize that, broadly speaking, the precision achieved are comparable to the 
	best results obtained in the Image CLEF 2011 conference~\cite{tsikrikaPK11}. 
	The conference overview only publishes the results for P@10 and P@15, which 
	are 0.632. and 0.510 respectively.  However, these results were achieved 
	combining textual and visual analysis techniques, using the three languages 
	in which the metadata of the documents are written (we only use English), 
	and exploiting feedback relevance techniques.  
	
	In 
	Figure~\ref{fig:CycleLenghtContribution}, we show the contribution of the 
	cycles depending on their length. We see that the cycles that contribute the 
	most are those with length equal to 2.  However, we count that, in average, 
	the query graphs only contain 1.56 cycles of this length.  This can be 
	caused either   because a) Wikipedia does not contain a large amount of such 
	cycles or because b) the cycles of length 2 are not always reliable, as 
	otherwise they would appear more frequently in the query graphs.  However, 
	according to our experiments, among all pairs of Wikipedia articles that are 
	connected, 11.47\% form a cycle of length 2, meaning that this structure is 
	not so infrequent. Then, we must assume the hypothesis that the cycles of 
	length 2 that contribute significantly to the quality of the results are 
	scarce. On the other hand, cycles of length 3, 4 and 5 contribute less, 
	compared to those of length 2, but appear more frequently, which encourages 
	us to focus on this type of structures.  

	\begin{table}[ht]
		\centering 
		\begin{tabular}{r|| c c c c c} 
			\hline\hline 
			Cycle Size  & Top 1 & Top 5 & Top 10 & Top 15\\  
			\hline 
			2&0.826&0.539&0.539&0.552\\
			3&0.833&0.578&0.519&0.513\\
			4&0.703&0.589&0.541&0.494\\
			5&0.788&0.624&0.588&0.547\\
			\hline 
		\end{tabular}
		\caption{Average precision of using expansion features of different 
			configurations of cycle lengths.} 
		\label{table:cycleResults} 
	\end{table}

	Regarding the observed proportion of articles and categories, in
	Figure~\ref{subfig:categoryRatio} we show the percentage of categories per
	cycle length. In general, we see that independently of the cycle length, a
  third of the nodes are categories, though there is a slightly increasing
	slope as the length increases. This fact, points out the importance of the
	category nodes, since those nodes are maintaining the cycles
	within a single or very related domain of knowledge. This observation may
	also justify the scarcity of cycles of length 2, given that cycles of this
	particular length are always composed by 2 articles, that may
	belong to very distant domains.

	Finally, in Figure~\ref{subfig:densityOfExtraEdges} we show the average
	density of extra edges with respect to the length of the cycles. We define
	the density of extra edges as the amount of edges minus the minimum required
	amount of edges to create a cycle divided by the maximum amount of
	possible edges of the cycle (remember that two consecutive nodes can be
  connected by two edges). From Figures~\ref{fig:CycleLenghtContribution} and
  ~\ref{subfig:densityOfExtraEdges}, we can see a correlation between denser cycles 
  and those that contribute more.

	\vspace{3mm} 
	Summarizing the characteristics 
	that let us differentiate good from bad cycles, we consent that:

	\begin{itemize}
		\item Cycles of length 2 are not reliable.
		\item Cycles of length 3, 4 and 5 are to be trusted to reach 
			articles that are strongly related with the input nodes.
		\item Around a third of the nodes of cycles have to be categories. 
			This ratio is expected to increase beyond the cycles of length 5.
		\item The expansion features obtained through the articles of dense 
			cycles are capable of leading to better results.
	\end{itemize}

	\subsection{Query Graph Expansion}

	
	Given the pipeline depicted in Figure~\ref{fig:pipeline}, at this point of 
	the process, the requests of the users are represented as a set of articles, 
	the input nodes. Given the input nodes, the goal of this module is to 
	select, among the Wikipedia nodes, those articles that allow constructing 
	query graphs that are similar to the ground truth ones, which were 
	previously described.

	\begin{figure}
		\centering
		\begin{subfigure}{0.4\linewidth}

			\includegraphics[width=\linewidth]{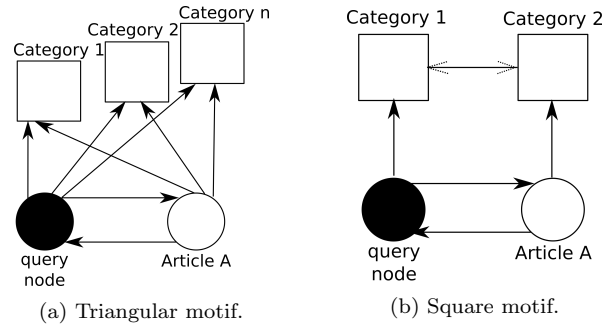}
			\caption{Triangular motif.}
			\label{fig:triangleShape} 
		\end{subfigure}
		\hspace{0.1\linewidth}
		\begin{subfigure}{0.4\linewidth}
			\includegraphics[width=\linewidth]{./images/squareShapeReduced.pdf}
			\caption{Square motif.}
			\label{fig:squareShape}
		\end{subfigure}
		\caption{Expansion motifs.}
	\end{figure}

	Based on these characteristics we propose the motifs depicted in 
	\hbox{Figures~\ref{fig:triangleShape}~and~\ref{fig:squareShape}}, which are 
	based on cycles of length 3 and 4 respectively. The motif  depicted in 
	Figure~\ref{fig:triangleShape} is called, from now on, \textbf{triangular 
	motif}, while the one depicted in Figure~\ref{fig:squareShape} is called 
	\textbf{square motif}. In the figures, the square nodes
	are categories, while round nodes are articles. The black round node is an
	input node,
	while the white round node is  article $A$, a new article selected as
	it forms a motif with the input node, and therefore, a node of the
	resulting query graph.  

	In the triangular motif we force the input node to be doubly linked with  
	article $A$. That means that the input node actually links, in Wikipedia, to 
	$A$, and $A$ links, reciprocally, to the input node.  Moreover, article $A$ 
	must belong to, at least, the same exact categories as the input node. This 
	structure guarantees a strong relation between the input node and 
	article $A$.  In the square motif of Figure~\ref{fig:squareShape}, the input 
	node and article $A$ must be also doubly linked. However, compared to 
	the triangular motif, it is just required that at least one of the 
	categories of the input node is inside one of the categories of $A$, or 
	\textit{vice versa} (depicted with a dashed arrow in  
	Figure~\ref{fig:squareShape}). This pattern still guarantees a strong 
	relation between the input node and article $A$, but is not as restrictive 
	as the one represented in Figure~\ref{fig:triangleShape}.  Both patterns are 
	chosen because they make sense from an intuitive point of view.  It is 
	expected that doubly linked articles that are also connected through 
	categories, are also close semantically (although the system has not done 
	any kind of semantic analysis), and the title of one can serve as 
	expansion feature of the other. Also because these cycles fulfill the edge 
	density and ratio of categories requirements.  
	To decide the length of the cycles that we base the motifs on, we ignore those of 
	length 2, as they resulted not to be trustful to identify strongly 
	related articles.  Larger cycles, as those with a length larger or equal to 
	5, have been also avoided for performance reasons. The traversal of larger cycles 
	expands too much the search space in Wikipedia, and would make it difficult 
	to identify them in a reasonable time for query expansion processes. Even 
	though performance is not the main goal of this piece of work, SQE has 
	always been designed with feasibility~in~mind. 

	The simplicity and efficiency of SQE consists in, given the input 
	nodes as a starting point, identify all the nodes of the Wikipedia graph 
	that are part of a motif and add them to the query graph. For each 
	article that we add in a query graph, we annotate the number of motifs 
	in which it has appeared. Note that the query graphs can be done using 
	triangular, square or using both types of motifs at the same time.  In 
	Section~\ref{sec:experiments} we analyze the benefits of using  of each 
	type of motif in different circumstances.  

	\begin{figure}[]
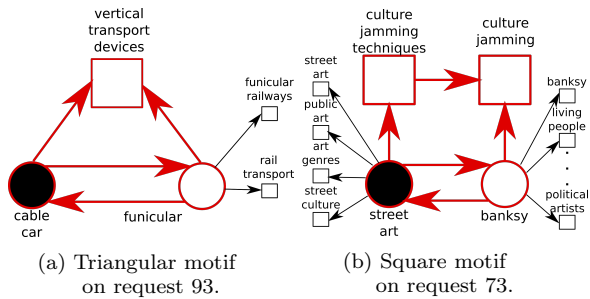

		\centering
		\begin{subfigure}{0.45\linewidth}
			\includegraphics[width=\linewidth]{./images/triangleShapeExample.pdf}
			\caption{\centering Triangular motif\newline on request 93.}
			\label{fig:triangleShapeEx} 
		\end{subfigure}
		\begin{subfigure}{0.45\linewidth}
			\includegraphics[width=\linewidth]{./images/squareShapeExample.pdf}
			\caption{\centering Square motif\newline on request 73.}
			\label{fig:squareShapeEx}
		\end{subfigure}
		\caption{Expansion motifs in action for Image CLEF requests.}
	\end{figure}

	In Figure~\ref{fig:triangleShapeEx} we show an example of a 
	triangular motif that adds an article to the query graph of the Image 
	CLEF request number 93, 
  whose input is ``cable cars''. Thanks to the motif, the article 
	\textbf{funicular}, that is a similar transport system, 
	becomes a part of the query graph. Similarly, in 
  Figure~\ref{fig:squareShapeEx}, for request number 73, whose 
	input is ``graffiti street art on walls'', the square motif introduces
	in its query graph the article \textbf{Banksy}, who is a 
	famous graffiti artist.

		\vspace{3mm}
	\subsection{Combining Query Graphs}

	According to the analysis presented in~\cite{guisado2015}, cycles of 
	different lengths may favor better precision for small amounts of results, 
	while other,  for larger amount of 
	results. Therefore, it is expected that, depending on the motif used to 
	create the query graph, the titles of its articles serve better as 
	expansion features for small or large tops of 
	retrieved	documents.

	\begin{figure}[]
		\centering
		\includegraphics[width=\linewidth]{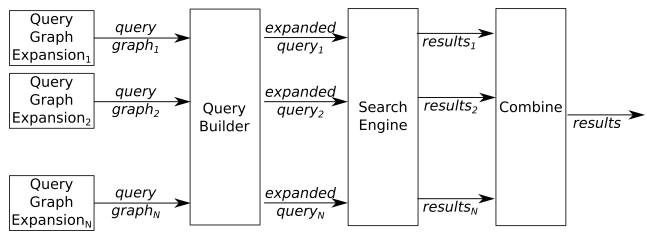}
		\caption{Combining several query graphs pipeline.}
		\label{fig:alternativePipeline}
	\end{figure}

	For that purpose we have implemented the pipeline depicted in 
	Figure~\ref{fig:alternativePipeline}. This pipeline allows configuring the 
	system in a way that, first, it creates several query graphs with different 
	configurations, for example one using the triangular motif, another using 
	the square motif and yet another one using a combination of both types of 
	motifs.  Then, using each of these query graphs, in combination with the 
  input and the entities (not depicted in the figure) the system creates a 
	query, as described in detail in Section~\ref{sec:experimentalSetup}. Then each 
	of these queries is run by the search engine which returns a set of 
  documents. Finally, the documents are combined into a single list of 
  documents that is returned to the user as the result to his/her request.

	In Section~\ref{sec:experiments}, we study the performance, in terms of 
	the quality of the results, but also in terms of the execution time of 
	this proposal and  also we discuss the proper way to combine the several 
	sets of results into a single one.

	\section{Experimental Setup} \label{sec:experimentalSetup}

	In this section we detail the whole system which is 
	necessary to better comprehend the experimental results. 

	{ \begin{center} \textbf{Entity Linker} \end{center}

	Entity linking  consists in identifying the entities of a text and link 
	them with a source of knowledge. In our case, it consists in identifying 
	the entities in the input, a set of keywords, and link them with 
	the input nodes. 
	Although entity linking is not the focus of this paper, we explain it 
	for the sake of clarity and due its importance for~the~overall~process. 

	The Entity Linker is based on two different libraries:
	Dexter~\cite{dexter2013} and   Alchemy\cite{alchemyWeb}. Dexter is an open
	source project that actually recognizes entities in a given text and links
	them with Wikipedia articles. On the other hand, Alchemy is a proprietary
	software which recognizes also entities but does not link them with
	Wikipedia. Both libraries have been tuned to extract entities from
	keyword-like texts, which are 2.4 terms long in 
	average~\cite{wolfram2001Jasist}.

	Since Dexter is capable of  recognizing the entities and link them
	with Wikipedia, which is the goal of this module, it is the first library
	that we use. Only if Dexter is not able to find any matching entry, we use
	Alchemy. In that case, once Alchemy returns the set of entities, we use
	Dexter again to process one by one and link them with Wikipedia. We have
	observed, that preprocessing the input with Alchemy and then using Dexter,
	allows Dexter to retrieve input nodes that it could not find with just the 
	input text.

	In Table~\ref{tab:entityLinking} we provide a few examples of inputs
	and the results of the entity linking process. We observe that for the 
	request number 73 and 93 of Image CLEF the process works properly and 
	returns two and one input nodes respectively, all of them referring to 
	correct entities within the input. However, for the input of the request 
	number 96, the process retrieves an input node which does not refer to 
	the correct entity of the input. Instead 
	of linking the entity shaking hands with the article $Handshake$, it 
	links it with the article $Shake$, which is a disambiguation page.  
	Finally, we see that the entity linker module its not able to retrieve 
	anything for the input of the request number 110. It is difficult for us 
	to understand the reasons behind these mistakes, since the entity 
	linking is carried on by Dexter. In case of not returning entities, the 
	pipeline is stopped and the system cannot look for expansion features 
	However, according to our experiments this is very rare.

	\begin{table}[]
		\centering
		\resizebox{\linewidth}{!}{%
			\begin{tabular}{l|l|l}
				Request ID            & Input                                         & Input Nodes           \\ \hline \hline
				\multirow{2}{*}{73} & \multirow{2}{*}{graffiti street art on walls} & Graffiti           \\
				&                                               & Street\_art        \\ \hline
				93					& cable car										& Cable\_car 			\\ \hline
				96					& shaking hands 								& Shake 			\\ \hline
				110                 & male color portrait                           &                    \\ \hline
			\end{tabular}
		}
		\caption{Entity linking on Image CLEF queries\protect\footnotemark.}
		\vspace{-3mm}
		\label{tab:entityLinking}
	\end{table} 

	\footnotetext{Real Wikipedia pages can be visited Wikipedia adding the prefix 
	https://en.wikipedia.org/wiki/ to the input node.}

	{\begin{center} \textbf{Query Builder \& Search Engine}\end{center}}
	Technically, we should differentiate between the request of the users, 
	in the form of a set of keywords, from the set of ``instructions'' that 
	the search engine is capable of understanding and processing. To the 
	latter we call it query. The goal of the query builder is precisely to 
	build a query that is actually understandable by the search engine.

	For this paper we have used Indri~\cite{strohman2005indri} as the 
	search engine that processes the query. Indri is a state of the art 
	open-source search engine that provides phrase matching, term proximity, 
	explicit phrase weighting and the usage of pseudo-relevance 
	techniques. Indri language consists of several operators that we 
	use to assemble the different elements that form the queries:

	\begin{itemize}
		\item \texttt{combine}: indicates that the different elements 
			have to be weighted equally and combined in an ``or'' 
			fashion within the search engine.
		\item \texttt{weight}: allows specifying a different weight for 
			each of the elements.
		\item \texttt{\#N}: ordered window -- elements must appear 
			ordered, with at most N-1 elements between each. 
			\texttt{\#1} indicates exact phrase matching.
	\end{itemize}

	At this point of the pipeline depicted in Figure~\ref{fig:pipeline}, the 
	intent of the user is described by: i) her/his input, which is written in 
	natural language, as a set of keywords, ii) the input nodes and iii) the 
	query graph. In order to build a query that Indri can understand, we treat  
	each of these components in a different way.

	The input keywords are assembled with the \texttt{combine} operator. To 
	process the input nodes we use their corresponding article titles.
	Each title is assembled with the phrase matching operator 
	\texttt{\#1}. Then, titles are assembled, 
	again, with the \texttt{combine} operator. In order to process the 
	query graph, we only use the articles  and  
	discard the categories. Articles are processed in the same way 
	as input nodes, via phrase matching, but they 
	are assembled together with the \texttt{weight} operator. The weight 
	of each title is the number of motifs in which it has appeared and 
  that we annotated during the query graph creation process.  Notice that this
  means that we 
	are also exploiting the structural properties to build the query. These 
	three elements can be used isolated or combined into a single query 
	by means of the \texttt{combine}~\hbox{operator}.

	\begin{query}[]
		\scriptsize
		\centering
		\fbox{\begin{minipage}{\dimexpr\linewidth-2\fboxsep-2\fboxrule\relax}
			{{\color{white}0}1:\color{white}\#ca}\#combine(\\
			{{\color{white}0}2:\color{white}\#ca}{\color{white}\#combine(}{\it \color{blue}\%input}\\
			{{\color{white}0}3:\color{white}\#ca}{\color{white}\#combine(}\#combine(graffiti street art on walls)\\
			{{\color{white}0}4:\color{white}\#ca}{\color{white}\#combine(}{\it \color{blue}\%entities}\\
			{{\color{white}0}5:\color{white}\#ca}{\color{white}\#combine(}\#combine(\#1(graffiti) \#1(street art))\\
			{{\color{white}0}6:\color{white}\#ca}{\color{white}\#combine(}{\it \color{blue}\%expansion features}\\
			{{\color{white}0}7:\color{white}\#ca}{\color{white}\#combine(}\#weight(\\
			{{\color{white}0}8:\color{white}\#ca}{\color{white}\#combine(\#weight(}5.0\#1(stencil){\color{white}aai}5.0\#1(yarn bombing)\\ 
			{{\color{white}0}9:\color{white}\#ca}{\color{white}\#combine(\#weight(}4.0\#1(above (artist)) 3.0\#1(banksy) \\
			{10:\color{white}\#ca}{\color{white}\#combine(\#weight(}3.0\#1(john fekner){\color{white}a}3.0\#1(urban art) \\ 
			{11:\color{white}\#ca}{\color{white}\#combine(\#weight(}3.0\#1(public art) \textbf{{\ldots}} \\
			{12:\color{white}\#ca}{\color{white}\#combine(\#weight})\\
			{13:\color{white}\#ca}{\color{white}\#combine})
		\end{minipage}}
		\caption{Expanded query written with the Indri language.}
		\label{queryExample}
	\end{query}

  Let us use a syntax sugar to refer to the queries that we may send to the 
	search engine. We call input, entities, and expansion features to 
	the Indri-formated input, titles of the input nodes and titles of the 
	articles of the query graph, respectively.  When we combine all these 
	components into a single query, we  talk about the expanded query.

	In Query~\ref{queryExample}, we show part of a query written in Indri 
  that combines the three components. The input is in the third line, the 
  entities are in the fifth line and the expansion features are from the 
	seventh line on. Note that the expansion features are weighted 
  according to what we explained, thus, \ttt{stencil} and \ttt{yarn bombing} 
	have appeared in five motifs, \ttt{above (artist)} in four and the rest 
	of the expansion features are the titles of articles that have appeared 
	in 3 motifs. In Section~\ref{sec:experiments} we analyze the best way to 
	combine those three components in order to achieve the best results.

	{\begin{center} \textbf{Datasets}\end{center}}
	{\flushleft For} the evaluation of SQE we use three datasets:

	{\flushleft \textbf{Image CLEF}}: This is a multimedia retrieval dataset.  
			The collection of results contains 237,434 images downloaded from 
			Wikipedia, which have short descriptions as metadata.  
			Approximately, 60\% of these descriptions contain texts in 
			English. The test collection also provides fifty requests. Each of 
			which consists of an input in the form of a set of keywords, a 
			brief description in natural language, and a set of relevant 
			images in the test collection. This dataset has been used during 
			the implementation of the query expansion system. 

			{\flushleft \textbf{CHiC 2012 \& CHiC 2013}}: These datasets are based on 
			cultural heritage retrieval. Both datasets shared the collection of 
			results, which contains 1,107,176 short documents. Each dataset 
      provides a set of fifty requests and their corresponding valid results. 
      These datasets  have 
			been only used for experimental reasons, and never during the 
			implementation of the system to avoid overfitting.

	{\begin{center} \textbf{Wikipedia Dump \& Graph Database Manager}\end{center}}
	We use the English Wikipedia dump of July 2nd, 2012 as our KB.  
	It has 9,483,031 articles and 99,675,360 links among articles,  
	1,320,671 categories, 3,795,869 links among categories and 41,490,074
	links among articles and categories.

	We use Sparksee~\cite{martinez2015Ideas} as our graph database manager. The 
	database requires about 8GB in disk. Notice that for each article we only 
	store its title because we do not need the content of the article to apply 
	SQE.

	{\begin{center} \textbf{Evaluation}\end{center}}
	To evaluate the results we use TrecEval, which is a program to evaluate 
	TREC results using the standard NIST evaluation procedure. This is 
	possible because CLEF datasets are TREC compatible. We focus on the 
	analysis of the system's precision for the top 5, 10, 15, 20, 30, 100, 
	200, 500 and 1000 results, which are the default tops in TrecEval.  
  Notice that the precision of the results is defined as the fraction of retrieved 
	documents that are relevant to the user.

	To show the statistical significance of SQE, with $p$<0,05, 
	we have done the paired t-test, which is used to compare two population 
	means, usually in 'before-after' studies. For the tests we have used as 
	the 'before' situation  the best configuration among the input, issued 
	by the user, the entities, or the combination of both the input and the 
	entities.
	\section{Experiments} \label{sec:experiments}

	\subsection{Query Graph Expansion Evaluation}

	We start analyzing the potential of the query graphs to improve the results.  
	For this reason, we have isolated the query graph expansion module from the 
	entity linker module, since the later could introduce errors that would 
	propagate and, hence, it would make it difficult to understand the real 
	benefit of using our proposed strategy. For that purpose, we have selected 
	manually the input nodes of the entities in the input.

	\begin{table}
		\centering
		\resizebox{\linewidth}{!}
		{
			\begin{tabular}{l|lllllllll|}
				\cline{2-10} & P@5   & P@10  & P @15 & P@20  & P@30 & P@100 & P@200  & P@500 & P@1000 \\ \hline
				\multicolumn{1}{|l|}{input}                                                      & 0.136 & 0.130 & 0.121 & 0.112 & 0.089  & 0.035 & 0.018  & 0.007 & 0.003  \\ \hline
				\multicolumn{1}{|l|}{entities}											     & 0.248 & 0.226 & 0.220 & 0.213 & 0.197  & 0.125 & 0.077  & 0.038 & 0.020  \\ \hline
				\multicolumn{1}{|l|}{\begin{tabular}[c]{@{}l@{}}input+\\entities\end{tabular}} & 0.244 & 0.220 & 0.213 & 0.210 & 0.195  & 0.127 & 0.081  & 0.040 & 0.021  \\ \hline \hline
				\multicolumn{1}{|l|}{\begin{tabular}[c]{@{}l@{}}expanded \\ query 1\end{tabular}}                                                     & 0.456\textdagger  & 0.402\textdagger  & 0.384\textdagger  & 0.349\textdagger  & 0.282\textdagger   & 0.147\textdagger  & 0.0859\textdagger  & 0.040\textdagger  & 0.020\textdagger   \\ \hline
				\multicolumn{1}{|l|}{\begin{tabular}[c]{@{}l@{}}expanded \\ query 2\end{tabular}}                                                     & 0.448\textdagger  & 0.414\textdagger  & 0.400\textdagger  & 0.379\textdagger  & 0.315\textdagger   & 0.171\textdagger  & 0.102\textdagger   & 0.048\textdagger  & 0.025\textdagger   \\ \hline
				\multicolumn{1}{|l|}{\begin{tabular}[c]{@{}l@{}}expanded \\ query 3\end{tabular}}                                                     & 0.444\textdagger  & 0.402\textdagger  & 0.387\textdagger  & 0.362\textdagger  & 0.301\textdagger   & 0.164\textdagger  & 0.104\textdagger   & 0.051\textdagger  & 0.027\textdagger    \\ \hline \hline
				\multicolumn{1}{|l|}{\begin{tabular}[c]{@{}l@{}}Upper bound\\ Cycles length 3\end{tabular}} & 0.578 & 0.519 & 0.513 & 0.415 & 0.339 & 0.163 & 0.095  & 0.042 & 0.023   \\ \hline
				\multicolumn{1}{|l|}{\begin{tabular}[c]{@{}l@{}}Upper bound\\ Cycles length 4\end{tabular}} & 0.589 & 0.541 & 0.494 & 0.485 & 0.382  & 0.188 & 0.117  & 0.054& 0.028   \\ \hline 
			\end{tabular}
		} 
		\caption{Precision obtained by different configurations at different 
		levels of precision. \textdagger~indicates statistically significant 
	improvement for the Image CLEF dataset.} 
	\vspace{-3mm} 
		\label{tab:ImageCLEFConfigurations}
	\end{table}

	For these experiments, we focus on the Image CLEF dataset, since it 
	is the dataset that was used in the analysis of~\cite{guisado2015}, and that 
	for which we have the ground truth. 
	As baselines we use the \texttt{input}, which is the text issued by 
	the user, the \ttt{entities}, whose input nodes have been manually 
	selected and the \texttt{input} + 
	\texttt{entities}.  Then, we study the results achieved using three 
	different ways of creating the expanded query (\ttt{input} + 
	\ttt{entities} + \ttt{expanded features}) that differ 
	in the way of creating the query graph, out of which the 
	expansion features are extracted.  In the \texttt{expanded query 1}, 
	the triangular motif is used, in the \texttt{expanded query 2} both 
	the triangular and the square motifs are used  to create the query 
	graph and in the \texttt{expanded query 3} only the square motif is 
	used. We also compare our current results with those obtained 
	in~\cite{guisado2015}, which we use as  an upper bound as they were 
	achieved using a ground truth query graph (\ttt{Upper bound Cycles length 3} 
	and \ttt{Upper bound Cycles length 4}). 
  
  In Table~\ref{tab:ImageCLEFConfigurations}, we see that the three
  query expansion configurations improve, with statistical significance, the
  precision achieved either for the input, the entities or the combination of
  both for all the tested levels of precision. This means that the achieved
  improvement is due to the introduction of the expansion features, and not only
  due to the identification and usage of the entities.

	We also see that the results achieved by the expanded queries 
	represent, in the worst case scenario (\ttt{expanded query 3}, P@20 
	- 0.362), the 71.41\% of the upper bound results (\ttt{cycles of 
	length 4}, P@20 - 0.485). In average, this percentage is 85.86\%, 
	which means that the proposed query expansion strategy is close to 
	the results achieved by the upper bound. 
	Note that the results achieved by the upper bound configurations are 
	due to the search of cycles within a controlled environment, which 
	are the ground truth query graphs created knowing  
	\textit{a priori} which were the valid documents for each user 
	request.
	Now, we are blindly creating the query graphs, traversing the whole 
	Wikipedia graph with the only information of the cycles 
	characteristics, out of which we have described the motifs.  So, it 
	is expected that the created query graphs and the ground truth query 
	graphs are not equal.  The former may lack some articles that existed 
	in the latter or it may have articles that~did~not~appear.  

	\begin{figure}
		\includegraphics[width=\linewidth]{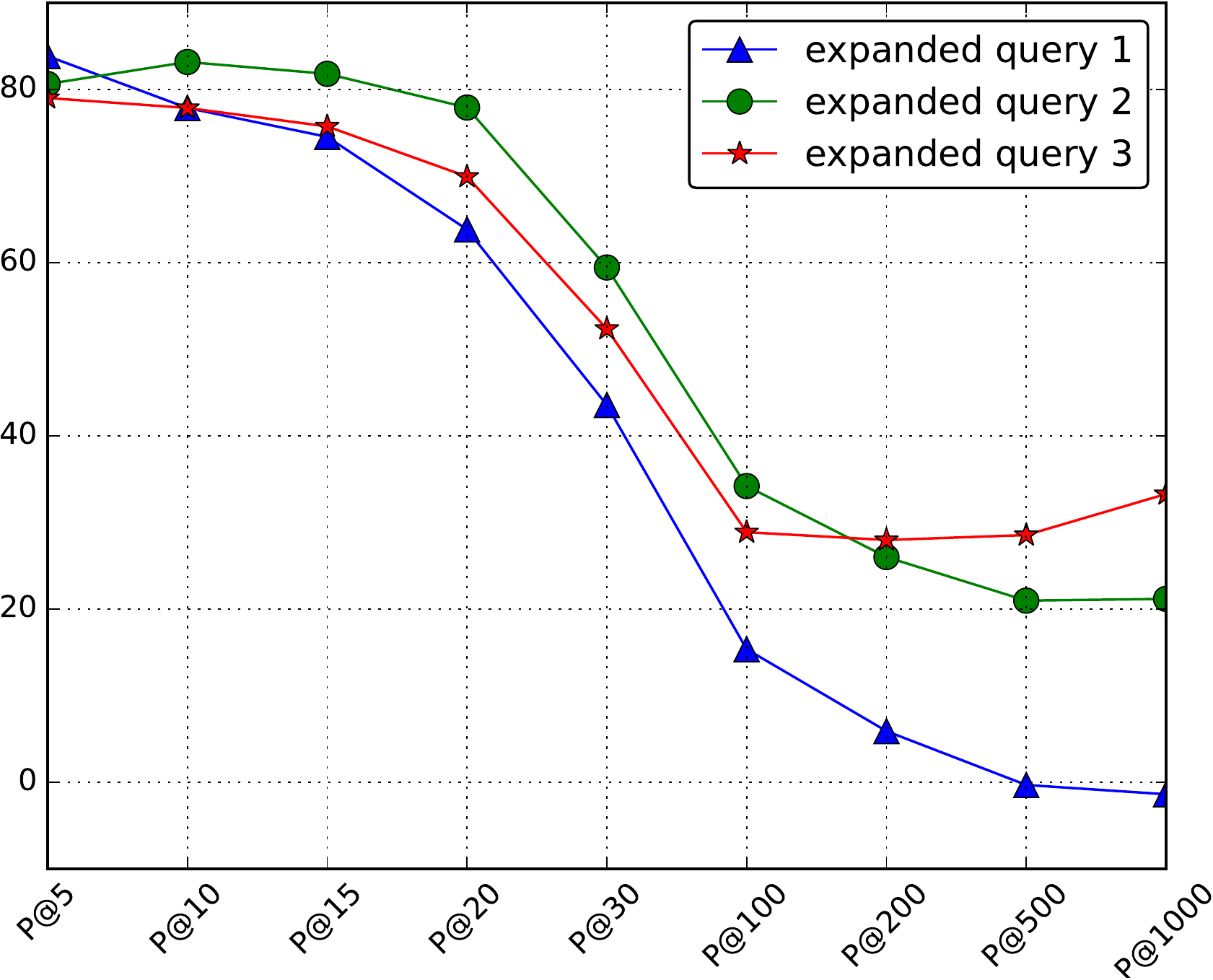}
		\caption{Percentual improvement over the maximum of input,  
		entities and input + entities.}
	\vspace{-3mm} 
		\label{fig:percentualImprovement}
	\end{figure}

	\begin{figure*}[]
		\centering
		\begin{subfigure}[b]{0.3\textwidth}
			\includegraphics[width=\linewidth]{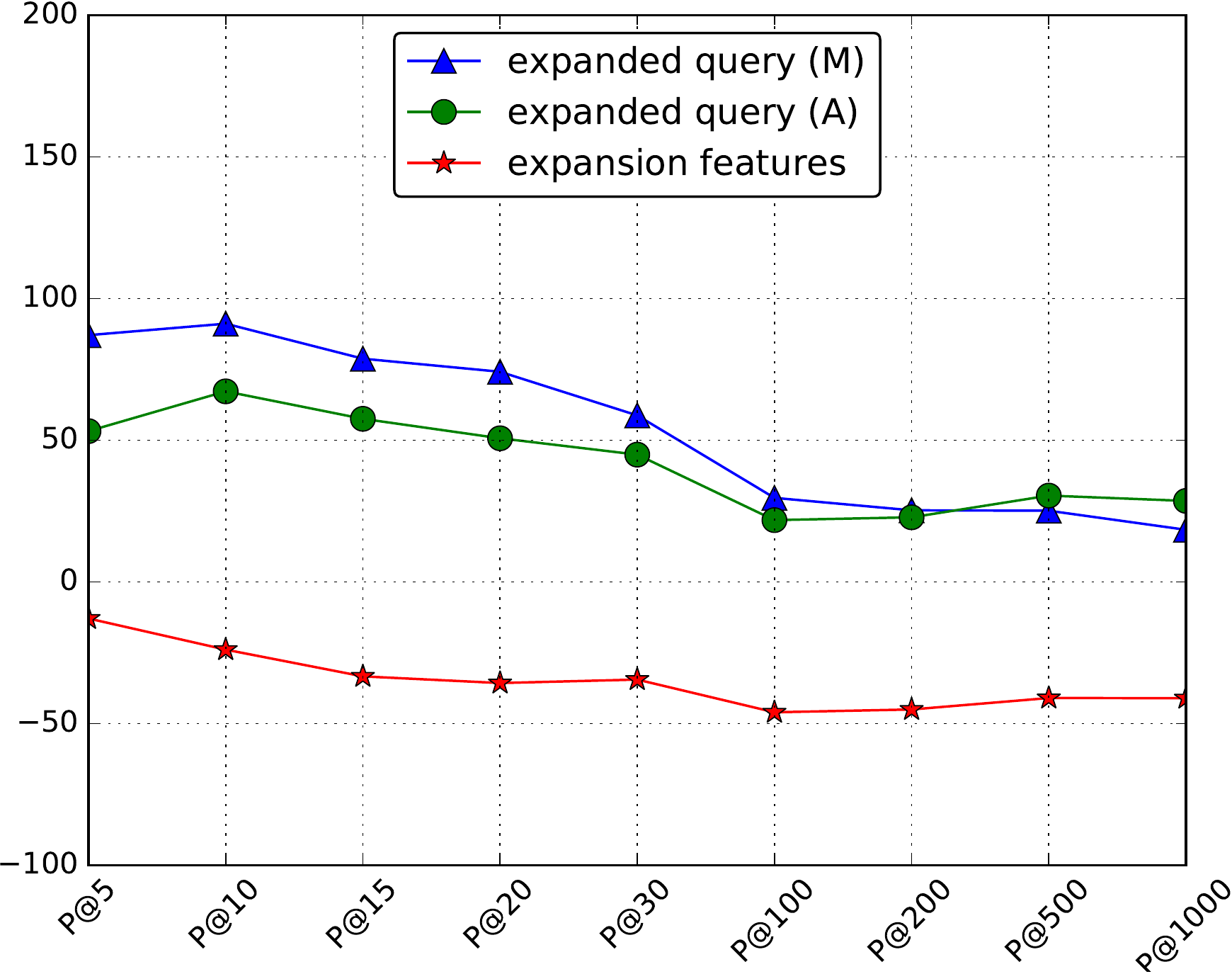}
			\caption{Image CLEF}
			\label{fig:imageClef} 
		\end{subfigure}
		\hfill
		\begin{subfigure}[b]{0.3\textwidth}
			\includegraphics[width=\linewidth]{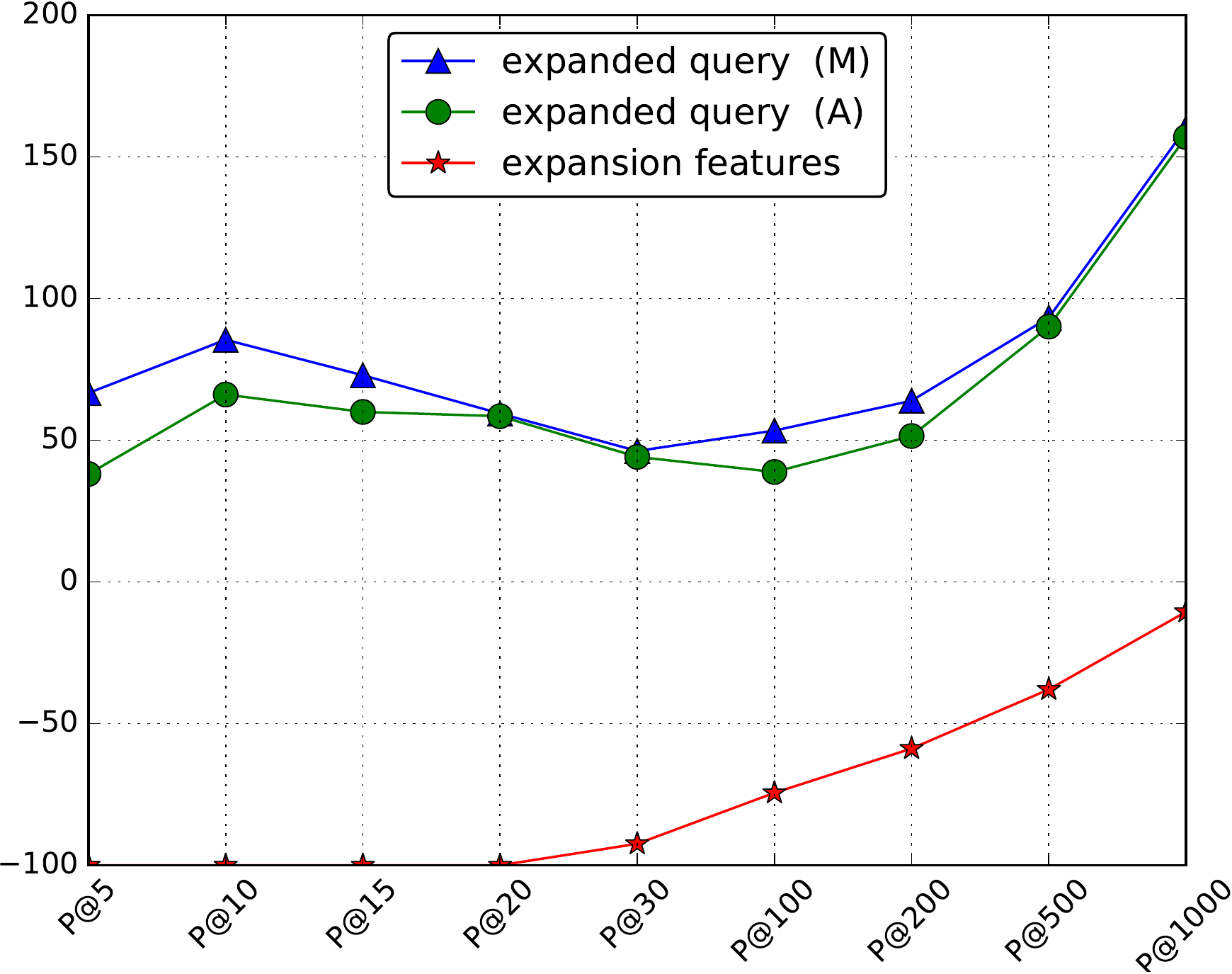}
			\caption{CHiC 2012}
			\label{fig:chic2012} 
		\end{subfigure}
		\hfill
		\begin{subfigure}[b]{0.3\textwidth}
			\includegraphics[width=\linewidth]{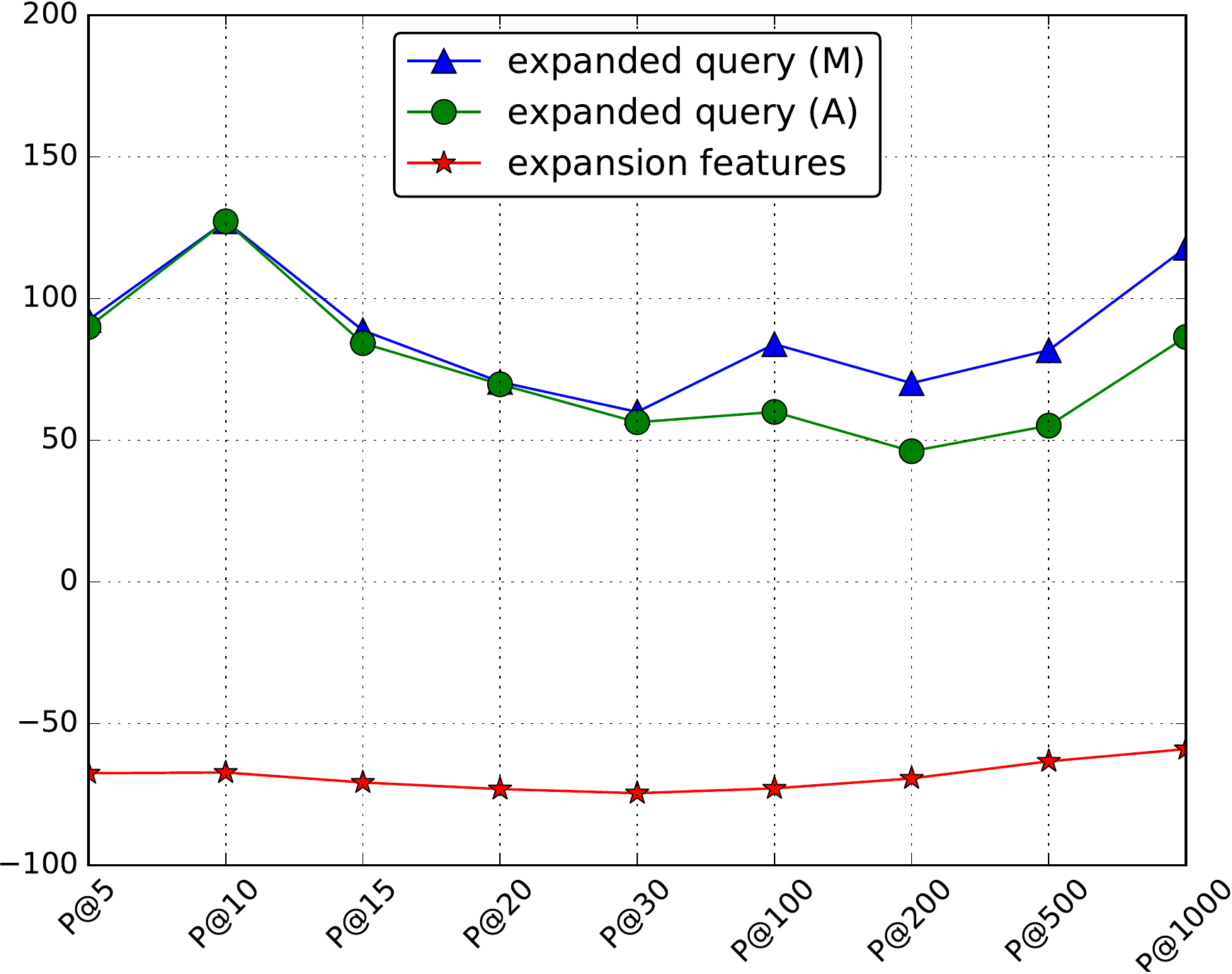}
			\caption{CHiC 2013}
			\label{fig:chic2013} 
		\end{subfigure}
		\caption{Percentual improvement of the query expansions 
			selecting the entities manually \texttt{expanded query (M)} and 
			automatically \texttt{expanded query (A)} and also of the expansion 
		features isolatedly.}
		\label{fig:percentual}
	\end{figure*}

	In Figure~\ref{fig:percentualImprovement} we show the percentual improvement 
	of the three expanded queries shown in  
	Table~\ref{tab:ImageCLEFConfigurations} with respect to the best result 
	achieved by the input, the entities or the combination of both. An analysis 
	of these configurations reveals three different ranges, depending on the 
	query expansion configuration that achieves the best results.  The first 
	range which includes the first five results, up to P@5, the second range 
	that goes from P@5 to P@100 and the third range from P@100 to P@1000. 

	{\flushleft}
	{\flushleft \textbf{Range P@1-P@5}}: The three configurations 
	achieve an improvement around the 80\%.  However, the one that 
	achieves the best results is the \texttt{expanded query 1}, whose 
	query graph is created only by means of triangular motifs. According 
	to our results, this configuration introduces 0,76 articles in the 
	query graph (and expansion features) per user request. This means 
	that this type of motif is very restrictive -- given an input node it 
	is difficult to find other articles that are related to it through 
	this type motif --, but very trustful. The introduction of the 
	expansion features via the triangular motifs allows the system to 
	achieve an improvement of 83.87\%. However, since there are just a 
	few expansion features, the expanded query is not very different 
	from the input issued by the user, and the improvement decreases 
	quickly as we  look at larger tops.

	{\flushleft \textbf{Range P@5-P@100}}: The best results are achieved 
	by the \texttt{expanded query 2} and the improvement goes from 
	83.85\% to 34.22\%. This configuration consists in 
	creating the query graphs using triangular and square motifs. The 
	combination of  both types of motifs introduces in average 20.96 
	expansion features per query. The introduction of the expansion 
	features that are obtained through the square motif, allow the 
	system to 
	introduce expansion features that are not as close to the original 
	input but still related and useful for larger tops. However, the 
  fact that these expansion features are combined with those introduced 
  by the triangular motifs makes this configuration the best for 
	this range in the middle between very small tops (P@5) and larger 
	ones (from P@100 to P@1000).

	{\flushleft \textbf{Range P@100-P@1000}}:	We observe that the 
	configuration that allows achieving the best results is the \texttt{expanded 
    query 3} which allows an improvement from 27.99\% to 33.30\%. 
	This configuration  introduces, in average, 20.48 expansion features 
  per query. The fact that these expansion features are not so tied to the 
	original input issued by the user enables to retrieve documents that were 
	not selected by the other configurations.  {\flushleft}

	To understand why the percentual improvement diminishes with the 
	size of the analyzed top, we need to understand that the average 
	number of correct documents per query is 68.8.  Hence, it is 
	difficult to keep improving when the amount of retrieved documents is  much 
	larger than the amount of actually valid documents.

	\subsection{System Evaluation}

	We now analyze the whole system including the entity linker module. 
	For that purpose, and according to the results previously shown, we 
	will use the variation of the system depicted in 
	Figure~\ref{fig:alternativePipeline}. We have configured the system 
	to use the set of results obtained by the configurations previously 
	discussed: \texttt{expanded query 1}, \texttt{expanded query 2} and 
	\texttt{expanded query  3} to build the list of results that are 
	returned to the user. Notice that, since the three expanded queries 
	are expressing the same user request but using different expansion features, 
	it is expected that the three sets of results share many documents. 
  As a consequence, in order to obtain the best of each 
	configuration, the ranges previously described have to be adjusted.  
	Experimentally, we have configured the system to build the final 
	list of results in a manner that the first 5 results come from 
  \texttt{expanded query 1}, then 30 results are included from 
	\ttt{expanded query 2}, and the rest of the results, up to 
	1000, come from \texttt{expanded query 3}.


	In Figure~\ref{fig:percentual} we 
	show the percentual improvement for Image CLEF, CHiC 2012 and CHiC 2013 
	over the best execution without the expansion features --using 
	only the \texttt{input}, only using the \texttt{entities} (manually and 
	automatically selected) or using both the \texttt{input} and the 
  \texttt{entities}--. More precisely, we show the percentual improvement  
	executing the expanded query, selecting manually the entities, 
	\texttt{expanded query (M)}, and also selecting the entities automatically, 
	as described in Section~\ref{sec:experimentalSetup}, \texttt{expanded query 
	(A)}. We also show the percentual improvement using only the 
  \texttt{expansion features} to retrieve the results. For the three datasets of
  Figure~\ref{fig:percentual}, we see that using the expansion features
  in an isolated way is not useful to improve the precision of the system, but
  diminishes the quality of the results.  That supports the idea of assembling
  the expanded query as described in 
	Section~\ref{sec:experimentalSetup} using i) the input issued by the users, 
	ii) the entities and iii) the expansion features. The input, even not being 
	the best way to express the real intention of the user, due to his/her lack 
	of knowledge and the vocabulary mismatch, is the only query form in which we 
	are sure that the system has not introduced any error and hence, it helps to 
	diminish errors that could be introduced later in the process. Introducing 
	the entities, reinforce the input, removing all signs of ambiguity from 
	the user intent. Finally, the expansion features introduce concepts that are 
	helpful to overcome the classical problems of information~retrieval. 

	Regarding the improvement achieved by SQE, which is depicted as 
	\texttt{expanded query (M)} and \texttt{expanded query (A)}, we observe that 
	it is  improving the results significantly. We also observe that there are 
	differences between selecting the entities manually or automatically. The 
	manual entity selection is almost an upper bound of SQE because it isolates 
	the creation of the query graphs from errors that could be introduced due to 
	the entity linking module,  as it is shown in Table~\ref{tab:entityLinking}.  
	Nonetheless, we observe that in the worst case scenario (Image CLEF, P@5), 
	the improvement achieved by \ttt{expanded query (A)} represents 81.89\% of 
	the result achieved by \ttt{expanded query~(M)}\footnote{This percentage can 
	be calculated using the results shown in 
	Table~\ref{tab:imageCLEFFinalResults}:  (0.380/0.248)/(0.464/0.248).}.  
	As shown in Figure~\ref{fig:chic2013}, there is also a difference 
	between the results achieved by \texttt{expanded query (M)} and 
	\texttt{expanded query (A)} for the larger tops, while in small tops is 
	imperceptible.  It is difficult to explain why in Image CLEF and CHiC 
	2012 the difference is noticeable in small tops, while in CHiC 2013 it is 
	noticeable  
	in larger tops. The most simple explanation is that since the set of 
	requests are different, it is difficult to expect the same behavior.  
	Another reason may be that, although similar efforts have been made to 
	select manually the entities, those of the CHiC 2013 dataset could not 
	be as precise as the ones in Image CLEF and CHiC 2012. Entity linking is 
	not the focus of this paper, however, improving the current entity 
	linking techniques used in our system would improve the results, making 
	it possible to achieve the results of selecting manually the entities 
	and the input nodes. 

	In 
	Tables~\ref{tab:imageCLEFFinalResults},~\ref{tab:chic2012FinalResults}~and~\ref{tab:chic2013FinalResults}, 
	we show the precision achieved in the three datasets. In particular, we show 
	the results achieved by our baselines, which are the \texttt{input}; the 
	entities selected manually (\texttt{entities~(M)}) or automatically 
	(\texttt{entities~(A)}); both the input and the entities (either manually or 
	automatically selected); and  only the \texttt{expansion features}. We also 
	show the results achieved by the expanded query, both selecting the entities 
	manually \texttt{expanded query~(M)} and automatically  \texttt{expanded 
	query~(A)}.  The results show that both \ttt{expanded query~(M)} and 
	\ttt{expanded query~(A)} present statistically significant  improvements  
	with respect to the baselines for the three datasets ($p$~<~0.05).

	\begin{table}[t]
		\centering
		\begin{subtable}{\linewidth}
			\resizebox{\linewidth}{!}
			{%
				\begin{tabular}{l|lllllllll|}
					\cline{2-10}
					& P@5   & P@10  & P @15 & P@20  & P\_@30 & P@100 & P@200 & P@500 & P@1000 \\ \hline
					\multicolumn{1}{|l|}{input}                                                                        & 0.136 & 0.130 & 0.121 & 0.112 & 0.089  & 0.035 & 0.018 & 0.007 & 0.003  \\ \hline
					\multicolumn{1}{|l|}{entities (M)}                                                           & 0.248 & 0.226 & 0.220 & 0.213 & 0.197  & 0.125 & 0.077 & 0.038 & 0.020  \\ \hline 
					\multicolumn{1}{|l|}{entities (A)}                                      					   & 0.156 & 0.134 & 0.145 & 0.147 & 0.137  & 0.107 & 0.069 & 0.035 & 0.022  \\\hline
					\multicolumn{1}{|l|}{\begin{tabular}[c]{@{}l@{}}input +\\ entities (M)\end{tabular}}         & 0.244 & 0.220 & 0.213 & 0.210 & 0.195  & 0.127 & 0.081 & 0.040 & 0.021  \\ \hline
					\multicolumn{1}{|l|}{\begin{tabular}[c]{@{}l@{}}input +\\ entities (A)\end{tabular}}       & 0.148 & 0.124 & 0.133 & 0.138 & 0.133  & 0.107 & 0.069 & 0.035 & 0.022  \\ \hline  \hline
					\multicolumn{1}{|l|}{\begin{tabular}[c]{@{}l@{}}expansion\\ features\end{tabular}}                 & 0.216 & 0.172 & 0.147 & 0.137 & 0.129  & 0.069 & 0.045 & 0.023 & 0.013  \\ \hline  \hline  
					\multicolumn{1}{|l|}{\begin{tabular}[c]{@{}l@{}} expanded\\ query (M)\\ \end{tabular}}           & 
					0.464\textdagger & 0.432\textdagger & 0.393\textdagger & 0.371\textdagger & 0.313\textdagger  & 0.165\textdagger & 0.102\textdagger & 0.050\textdagger & 0.027\textdagger  \\\hline  
					\multicolumn{1}{|l|}{\begin{tabular}[c]{@{}l@{}} expanded\\ query (A)\\ \end{tabular}} 		   & 
					0.380\textdagger & 0.378\textdagger & 0.347\textdagger & 0.321\textdagger & 0.286\textdagger  & 0.155\textdagger & 0.100\textdagger & 0.052\textdagger & 0.029\textdagger  \\ \hline
				\end{tabular}
			}
			\caption{Image CLEF results.}
			\label{tab:imageCLEFFinalResults} 
			\vspace{.5cm}
		\end{subtable}
		\begin{subtable} {\linewidth} 
			\resizebox{\linewidth}{!}{%
				\begin{tabular}{l|lllllllll|}
					\cline{2-10}
					& P@5   & P@10  & P @15 & P@20  & P\_@30 & P@100 & P@200 & P@500 & P@1000 \\ \hline
					\multicolumn{1}{|l|}{input}                                                                        & 0.148 & 0.100 & 0.084 & 0.077 & 0.074  & 0.034 & 0.018 & 0.007 & 0.004  \\ \hline
					\multicolumn{1}{|l|}{entities (M)}                                                            & 0.156 & 0.118 & 0.108 & 0.101 & 0.093  & 0.042 & 0.023 & 0.010 & 0.005  \\ \hline
					\multicolumn{1}{|l|}{entities (A)}                  										   & 0.100 & 0.072 & 0.067 & 0.061 & 0.053  & 0.021 & 0.011 & 0.007 & 0.004  \\ \hline
					\multicolumn{1}{|l|}{\begin{tabular}[c]{@{}l@{}}input +\\ entities (M)\end{tabular}}         & 0.168 & 0.124 & 0.113 & 0.106 & 0.097  & 0.044 & 0.023 & 0.010 & 0.005  \\ \hline
					\multicolumn{1}{|l|}{\begin{tabular}[c]{@{}l@{}}input +\\ entities (A)\end{tabular}}        & 0.116 & 0.086 & 0.076 & 0.068 & 0.057  & 0.022 & 0.012 & 0.007 & 0.004  \\ \hline   \hline 
					\multicolumn{1}{|l|}{\begin{tabular}[c]{@{}l@{}}expansion\\ features\end{tabular}}                 & 0.000 & 0.000 & 0.000 & 0.000 & 0.007  & 0.011 & 0.010 & 0.006 & 0.005  \\ \hline           \hline 
					\multicolumn{1}{|l|}{\begin{tabular}[c]{@{}l@{}} expanded\\ query (M)\end{tabular}}             & 0.280\textdagger  & 0.230\textdagger  & 0.196\textdagger  & 0.169\textdagger  & 0.141\textdagger   & 0.067\textdagger  & 0.038\textdagger  & 0.020\textdagger  & 0.013\textdagger   \\ \hline
					\multicolumn{1}{|l|}{\begin{tabular}[c]{@{}l@{}} expanded\\ query (A)\end{tabular}} 			   & 0.232\textdagger  & 0.206\textdagger  & 0.181\textdagger  & 0.168\textdagger  & 0.139\textdagger   & 0.061\textdagger  & 0.035\textdagger  & 0.019\textdagger  & 0.013\textdagger   \\ \hline
				\end{tabular}
			}
			\caption{CHiC 2012 results.}
			\label{tab:chic2012FinalResults}
			\vspace{.5cm}
		\end{subtable}
		\begin{subtable}{\linewidth} 
			\resizebox{\linewidth}{!}{%
				\begin{tabular}{l|lllllllll|}
					\cline{2-10}
					& P@5   & P@10  & P @15 & P@20  & P\_@30 & P@100 & P@200 & P@500 & P@1000 \\ \hline
					\multicolumn{1}{|l|}{input}                                                                        & 0.160 & 0.110 & 0.101 & 0.092 & 0.084  & 0.045 & 0.028 & 0.011 & 0.006  \\ \hline
					\multicolumn{1}{|l|}{entities (M)}                                                           & 0.132 & 0.110 & 0.119 & 0.115 & 0.104  & 0.054 & 0.035 & 0.016 & 0.009  \\  \hline
					\multicolumn{1}{|l|}{entities (A)}															   & 0.104 & 0.078 & 0.076 & 0.065 & 0.058  & 0.034 & 0.026 & 0.015 & 0.008  \\ \hline
					\multicolumn{1}{|l|}{\begin{tabular}[c]{@{}l@{}} 
						input +\\entities (M) \end{tabular}}        & 0.132 & 0.110 & 0.119 & 0.119 & 0.110  & 0.056 & 0.036 & 0.017 & 0.009  \\ \hline
					\multicolumn{1}{|l|}{\begin{tabular}[c]{@{}l@{}}input +\\ entities (A) \end{tabular}}  	   & 0.104 & 0.082 & 0.080 & 0.071 & 0.062  & 0.035 & 0.026 & 0.015 & 0.008  \\ \hline   \hline 
					\multicolumn{1}{|l|}{\begin{tabular}[c]{@{}l@{}}expansion\\ features\end{tabular}}                 & 0.052 & 0.036 & 0.035 & 0.032 & 0.028  & 0.015 & 0.011 & 0.006 & 0.004  \\ \hline   \hline 
					\multicolumn{1}{|l|}{\begin{tabular}[c]{@{}l@{}}expanded\\ query (M) \end{tabular}}              & 0.308\textdagger  & 0.250\textdagger  & 0.224\textdagger  & 0.203\textdagger  & 0.176\textdagger   & 0.103\textdagger  & 0.062\textdagger  & 0.030 & 0.020\textdagger   \\ \hline
					\multicolumn{1}{|l|}{\begin{tabular}[c]{@{}l@{}}expanded\\ query (A) \end{tabular}} 			   & 0.304\textdagger  & 0.250\textdagger  & 0.219\textdagger  & 0.202\textdagger  & 0.172\textdagger   & 0.090\textdagger  & 0.053\textdagger  & 0.026\textdagger  & 0.017\textdagger   \\ \hline
				\end{tabular}
			}
			\caption{CHiC 2013 results.}
			\label{tab:chic2013FinalResults}
		\end{subtable}
		\vspace{-3mm}
		\caption{Comparison of the precision achieved in the 
			datasets. \textdagger~indicates statistically significant 
			improvement. 
		\vspace{-5mm}
		} 
		\label{tab:precisionResults}
	\end{table}

  In Table~\ref{tab:imageCLEFFinalResults}, we show that the combination of query 
  graphs allows achieving even better results than each of the configurations 
  independently in their best range, as shown in Table~\ref{tab:ImageCLEFConfigurations}.
	This supports our strategy of combining the results 
	obtained by different expanded queries to 
	improve the quality of the results independently the amount of 
	them.

	Focusing only in \texttt{expanded query (A)} we also observe 
	differences among the results for the three datasets. At first sight we could think 
	that it performs better with Image CLEF because the precision 
	achieved with this dataset goes from 0.380 (P@5) to 0.0029(P@1000), 
	while for CHiC 2012 and 2013 it goes from 0.232 to 0.013 and from 
	0.304 to 0.017 respectively. We could also think this could be due 
	to an overfitting of SQE for Image CLEF, since it is the 
	dataset that has been used during the development of the system. However, there 
	are objective facts that explain this behavior. First, the document 
	collection of Image CLEF consists of 237,434 documents, while the document 
	collection of the CHiC datasets has 
	1,107,176. This makes Image CLEF an easier dataset. Moreover, Image CLEF  has an 
	average of 68.8 correct results per request, while CHiC 2012 and 
	CHiC 2013 have 31.32 and 50.6 respectively. In addition, all the 
	requests in Image CLEF have at least 1 correct result, while in CHiC 
	2012 there are 14 requests (out of 50) that do not have any correct results 
	and in CHiC 2013 there is 1 request without any correct result.  
	From an absolute value perspective, it is easier for the system to 
	achieve good results in terms of precision, when most of the 
	requests have valid results, and even easier if the number of results 
	is higher.
	Hence, Image CLEF is the one that achieves better results, CHiC 2013 
	comes next and finally, CHiC 2102. Moreover, although from an 
	absolute value point of view (as show in 
	Table~\ref{tab:precisionResults}) 
	it may appear that our system works better for Image CLEF, from a 
	relative point of view (as depicted in  
	Figure~\ref{fig:percentual}) 
	we see that the percentual improvement for the three datasets is 
	equivalent, and even better for the CHiC 2013 dataset.

	\begin{figure}[t]
		\includegraphics[width=\linewidth]{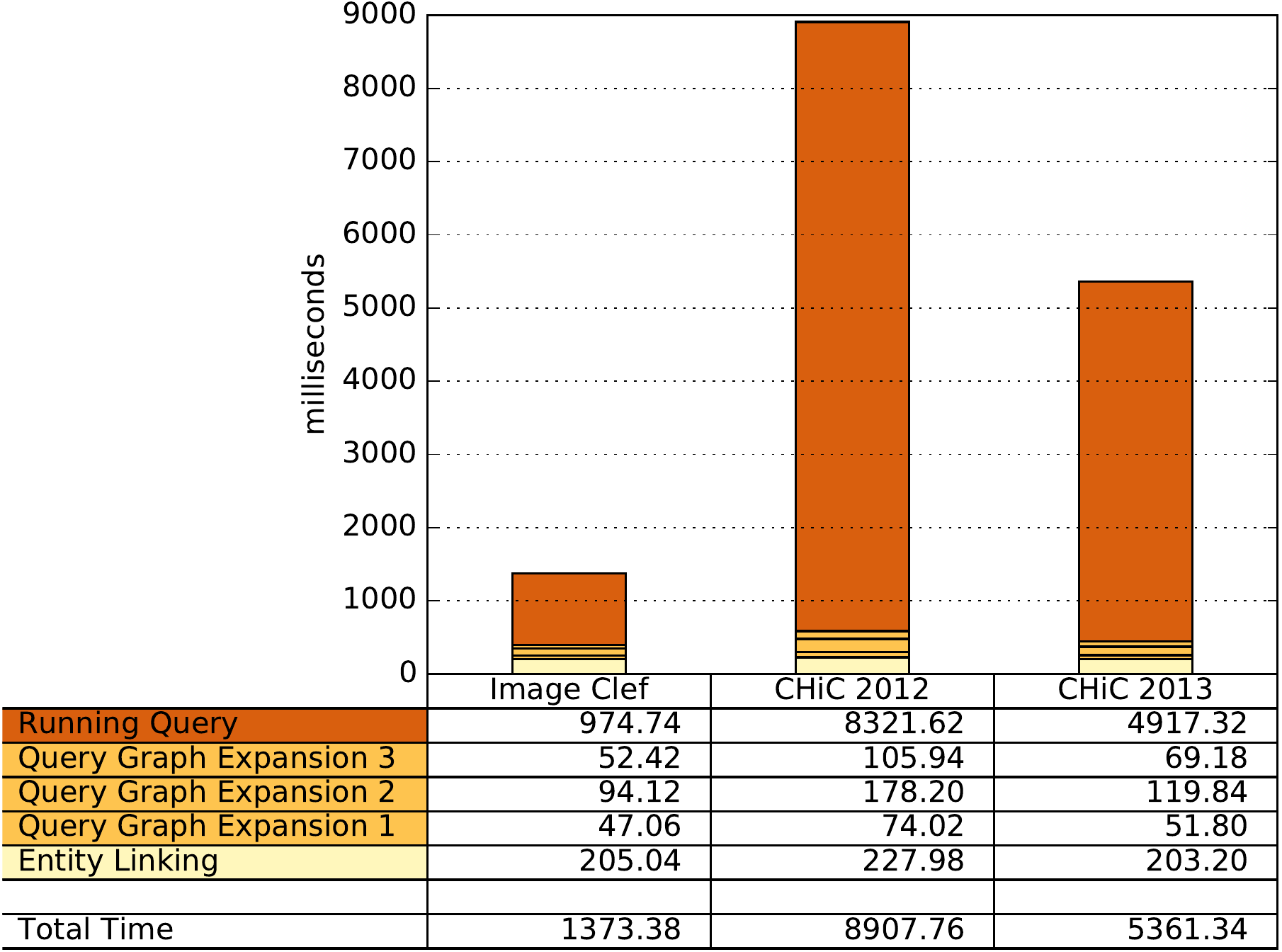}
	\vspace{-7mm} 
		\caption{Execution times in milliseconds.} \label{fig:executionTime} 
	\vspace{-3mm} 
	\end{figure}

	\subsection{System Performance}

	We also evaluate the performance of SQE. Even though it
	is not the main concern of this paper to address possible bottlenecks that
	might prevent SQE to be applied in a practical context, we show
	that it incurs in a negligible overhead. Note that we have not used any  
	technique, such as indexing or exploiting parallelism, to speed up the 
	process.


	In Figure~\ref{fig:executionTime}, we show the average  execution time per 
	request for each of the three datasets that have been used in this paper, 
	assuming that the entities are selected automatically (\texttt{expanded 
	query (A)}). We divide the execution time into three parts. The \ttt{Entity 
	Linking} time, which is the time that Dexter and Alchemy need to do the 
	entity linking. The query graph expansion time, which  is  depicted as 
	\texttt{Query Graph Expansion 1, 2, 3} which corresponds to the time spent 
	creating the query graphs of the expanded query 1, 2, 3, previously 
	described.
	And the \ttt{Running Query} time which is the time spent running the query 
	by the search engine. According to the results, the average time to run a 
	request and to obtain the documents goes from 1373.38 ms. (Image CLEF) to 
	8907.76 ms. (CHiC 2012) which is still far from being considered ``real 
	time''. Nonetheless, notice that the total query expansion time is 
	negligible compared with the whole process. In the worst case scenario, 
	which is the Image CLEF dataset, the expansion time represents 14\% of the 
	running time, while in the two other datasets this only represents 4\%.  
	Moreover, this time would probably be easily reduced by parallelizing the 
	expansion process,  which would reduce the expansion process time to the 
	maximum of the expansions times, instead of the aggregation. That could also 
	be done for the running time. Regarding the differences among the running 
	time of the three datasets, it is difficult for us to justify them or even 
	explain them, because the process is carried on by Indri. We do not blame in 
	any case Indri for the running query time, because it could be probably 
	reduced indexing the document collection in a better way. However, we have 
	detected, and it is expected to happen, that the more expansion features 
	introduced during the query expansion, the longer it takes to solve 
	the query. According to our results the average number of expansion features 
	for the queries of Image CLEF, CHiC 2012 and CHiC 2013 are 26.7, 46,06 and 
	33,52 respectively, which is correlated with the time spent running the 
	query.

	\begin{table}[]
		\centering
		\begin{subtable}{\linewidth}
			\resizebox{\linewidth}{!}
			{
				\begin{tabular}{l|r|r||r|r||r|r||r|r||r|r|}
					\cline{2-11}
					& \multicolumn{1}{c|}{P@5} & \multicolumn{1}{c||}{GAIN} & \multicolumn{1}{c|}{P@10} & \multicolumn{1}{c||}{GAIN} & \multicolumn{1}{c|}{P@15} & \multicolumn{1}{c||}{GAIN} & \multicolumn{1}{c|}{P@20} & \multicolumn{1}{c||}{GAIN} & \multicolumn{1}{c|}{P@30} & \multicolumn{1}{c|}{GAIN} \\ \hline
					\multicolumn{1}{|l|}{input }                                                                 & 0.000                    & -100.00\%                 & 0.000                     & -100.00\%                 & 0.000                     & -100.00\%                 & 0.001                     & -99.11\%                  & 0.000                     & -99.22\%                  \\ \hline
					\multicolumn{1}{|l|}{\begin{tabular}[c]{@{}l@{}}entities (A)\\  \end{tabular}}         & 0.004                    & -97.44\%                  & 0.004                     & -97.01\%                  & 0.004                     & -97.25\%                  & 0.003                     & -97.96\%                  & 0.002                     & -98.54\%                  \\ \hline
					\multicolumn{1}{|l|}{\begin{tabular}[c]{@{}l@{}}input  +\\ entities (A)\\ \end{tabular}} & 0.004                    & -97.30\%                  & 0.002                     & -98.39\%                  & 0.003                     & -97.98\%                  & 0.003                     & -97.83\%                  & 0.003                     & -97.96\%                  \\ \hline
					\multicolumn{1}{|l|}{\begin{tabular}[c]{@{}l@{}}expanded 
						\\query (A) \end{tabular}}        & 0.432                    & \textbf{+13.68\%}         & 0.37                      & -2.12\%                   & 0.348                     & \textbf{+0.39\%}          & 0.323                     & \textbf{+0.62\%}          & 0.289                     & \textbf{+1.15\%}          \\ \hline
				\end{tabular}
			}
			\caption{Image CLEF results.}
			\vspace{.25cm}
			\label{tab:imagePSRF}
		\end{subtable}
		\begin{subtable}{\linewidth}
			\resizebox{\linewidth}{!}
			{

				\begin{tabular}{l|r|r||r|r||r|r||r|r||r|r|}
					\cline{2-11}
					& \multicolumn{1}{c|}{P@5} & \multicolumn{1}{c||}{GAIN} & \multicolumn{1}{c|}{P@10} & \multicolumn{1}{c||}{GAIN} & \multicolumn{1}{c|}{P@15} & \multicolumn{1}{c||}{GAIN} & \multicolumn{1}{c|}{P@20} & \multicolumn{1}{c||}{GAIN} & \multicolumn{1}{c|}{P@30} & \multicolumn{1}{c|}{GAIN} \\ \hline
					\multicolumn{1}{|l|}{input }                                                                    
					& 0.000     & -100.00\%       & 0.002 & -98.00\%        & 0.0013 & -98.45\%        & 0.001 & -98.70\%        & 0.0007 & -99.05\%        \\ \hline
					\multicolumn{1}{|l|}{\begin{tabular}[c]{@{}l@{}}entities 
						(A) \\ \end{tabular}}            & 0.000     & -100.00\%       & 0.008 & -88.89\%        & 0.0053 & -92.05\%        & 0.005 & -91.80\%        & 0.0047 & -91.08\%        \\ \hline
					\multicolumn{1}{|l|}{\begin{tabular}[c]{@{}l@{}}input 
						+\\  entities (A) \\ \end{tabular}} & 0.000     & -100.00\%       & 0.004 & -95.35\%        & 0.0027 & -96.45\%        & 0.002 & -97.06\%        & 0.0013 & -97.73\%        \\\hline

					\multicolumn{1}{|l|}{\begin{tabular}[c]{@{}l@{}}expanded\\ query  (A) \end{tabular}}           & 0.244 & \textbf{5.17\%} & 0.218 & \textbf{5.83\%} & 0.1933 & \textbf{6.60\%} & 0.173 & \textbf{2.98\%} & 0.1447 & \textbf{3.85\%} \\ \hline
				\end{tabular}
			}
			\caption{CHiC 2012 results.}
			\label{tab:chic12PSRF}
			\vspace{.25cm}
		\end{subtable}
		\begin{subtable}{\linewidth}
			\resizebox{\linewidth}{!}
			{
				\begin{tabular}{l|r|r||r|r||r|r||r|r||r|r|}
					\cline{2-11}
					& \multicolumn{1}{c|}{P@5} & \multicolumn{1}{c||}{GAIN} & \multicolumn{1}{c|}{P@10} & \multicolumn{1}{c||}{GAIN} & \multicolumn{1}{c|}{P@15} & \multicolumn{1}{c||}{GAIN} & \multicolumn{1}{c|}{P@20} & \multicolumn{1}{c||}{GAIN} & \multicolumn{1}{c|}{P@30} & \multicolumn{1}{c|}{GAIN} \\ \hline

					\multicolumn{1}{|l|}{input }                                                                   & 0.000     & -100.00\% & 0.004 & -96.36\%        & 0.004  & -96.05\%        & 0.003 & -96.74\%        & 0.0033 & -96.07\%         \\ \hline
					\multicolumn{1}{|l|}{\begin{tabular}[c]{@{}l@{}}entities (A) \\ \end{tabular}}           & 0.000     & -100.00\% & 0.008 & -89.74\%        & 0.0067 & -91.18\%        & 0.008 & -87.69\%        & 0.0067 & -88.45\%         \\ \hline
					\multicolumn{1}{|l|}{\begin{tabular}[c]{@{}l@{}}input + \\ entities (A)\\ \end{tabular}} & 0.004 & -96.15\%  & 0.006 & -92.68\%        & 0.0053 & -93.38\%        & 0.006 & -91.55\%        & 0.0053 & -91.45\%         \\ \hline
					\multicolumn{1}{|l|}{\begin{tabular}[c]{@{}l@{}}expanded 
						\\query (A) \end{tabular}}           & 0.288 & -5.26\%   & 0.264 & \textbf{5.60\%} & 0.2373 & \textbf{8.52\%} & 0.22  & \textbf{8.91\%} & 0.1933 & \textbf{12.39\%} \\ \hline
				\end{tabular}
			}
			\caption{CHiC 2013 results.}
			\vspace{.3cm}
			\label{tab:chic13PSRF}
		\end{subtable}
		\vspace{-5mm}
		\caption{Comparison of the precision achieved in the datasets using 
		pseudo-relevance feedback techniques.}
		\vspace{-3mm}
	\end{table}
	
	\subsection{Pseudo-Relevance Feedback benefits}
  We have implemented a query expansion technique that is orthogonal to many 
	other techniques, some of them reviewed in Section~\ref{sec:relatedWork}.  
  As an example, we have combined SQE with Pseudo-relevance 
	feedback, another query expansion approach. Pseudo-relevance feedback 
	techniques consist in assuming that the initial set of retrieved results are 
	good and, hence, they can be used as a source of reliable expansion 
	features. In 
	Tables~\ref{tab:imagePSRF},~\ref{tab:chic12PSRF}~and~\ref{tab:chic13PSRF} we 
	show the results achieved by combining the \texttt{input}, the 
	\texttt{entities (A)}, and the \texttt{expanded query~(A)} with 
	pseudo-relevance feedback.  We have used the implementation of 
  pseudo-relevance that is inbuilt in Indri and we have used the default 
  configuration.  For Image CLEF,  we observe an important 
	improvement for the precision of P@5, up to 13.68\%, while the rest of the 
	improvements are not as remarkable. For the CHiC datasets the improvement is 
	more homogeneous for most of the observed tops.  Moreover, notice that pseudo-relevance 
	feedback techniques are not capable of improving the results of non-expanded 
	queries, on the contrary they tend to diminish the quality of the results.  
	We are not familiar with relevance feedback techniques nor their current 
	implementation in Indri, thus we have not tuned the build-in pseudo-relevance feedback 
	of Indri to better fit with our way of expressing the expanded query, which 
  would probably turn up in better results. Nonetheless, our results show the 
  potential of SQE in combination with other techniques.

	\section{Related Work}\label{sec:relatedWork}
    
	Query expansion techniques can be classified into several families depending 
	on the methods used to obtain the expansion features:  
	linguistic analysis~\cite{Paice1994}, query 
	specific~\cite{DBLP:journals/ipm/ChangOK06}, query-log 
	analysis~\cite{Song2012}, and external source of information such as KB. We 
	focus on the KBs family.


	The classical use of KBs consists in analyzing semantically their corpus  to 
	identify relevant information for the query that better describe the users 
	request. Particularly, Wikipedia has become a frequently used large corpus of 
	information.  For example, Egozi et al.~\cite{egozi2011} present an 
	interesting technique for query rewriting based on explicit semantic 
	analysis, where they postprocess queries obtained from pseudo-relevance 
	feedback using a KB. This technique depends on the quality of the 
	pseudo-relevance feedback expansion, which is very poor in our document 
	collection unless we previously expand the query, as seen in 
	Section~\ref{sec:experiments}. In~\cite{de2015knowledge}, the expansion 
	features are extracted out of the most important terms of the Wikipedia 
	article and are calculated with classical TF-IDF.  The corpus is also 
	used to derive search support tools. For instance, 
	in~\cite{hu2009understanding}, Wikipedia is used to build a map of concepts, 
	then, user requests are mapped onto those concepts, which make it easier for 
	the search engine to resolve them.  In~\cite{gan2015improving}, the textual 
	corpus of Wikipedia is used to derive a Markov network  to improve query 
	expansion.  There is a family of techniques that consist in using the anchor 
	texts of the links of the KB to identify the expansion features.  
	In~\cite{Eiron2003}, the authors show that anchor texts are similar to real 
	queries regarding to term distribution and length, therefore they can be a 
	source of expansion features. Also, in~\cite{dang2010query}, they are 
	exploited to build a virtual query log, that can be used to reformulate the 
	queries.   In~\cite{Arguello2008}, the authors propose a query expansion 
	method for blog recommendation. Their method is based on the analysis of 
	links. The anchor text of the most important twenty links is used to expand the 
	query which results in a significant improvement in terms of precision.  
	Such an approach could be used in our work to rate the importance of the 
	links, and then, include the strength of connections in the motifs.  In 
	contrast to previous works, in this paper we do not focus on analyzing the 
	content of the KB, or ranking the links among articles using other 
	techniques. We rely exclusively on the structure of the KB.  However, SQE 
	could be combined with most of the techniques that exploit the content of 
	the KB.  
	
	\section{Conclusions and Future Work} \label{sec:conclusions}

	In this paper we  contribute in opening a new research direction in the 
	field of query expansion. We have 
	proposed SQE as a query expansion technique that relies on the 
	underlying network structures of KBs, which, to the best of our knowledge, 
	have been barely exploited so far. From the analysis of a KB structure we have 
	defined a set of structural motifs which allow relating their tightly linked 
	entries.

	Given a user request, which is the input of the system, usually 
	represented as a set of keywords, SQE identifies its entities and links them 
	with the corresponding nodes of the KB, which we have named input nodes. 
	Then, only from the structural motifs and with no need of semantic 
	analysis, SQE relates the input nodes with a set of semantically connected 
	entries, out of which the expansion features are extracted.  

	There are many expansion techniques, which differ in the way they identify 
	the expansion features. Some of those techniques use query logs to rewrite 
	the queries, based on previous users behavior, others apply complex 
	semantic analysis, so the system can understand the real intent of the 
	user. There are many others that are not cited to synthesize.  SQE does not 
	expect to substitute them, we propose it as a query expansion method that 
	relies on handy external KBs for those systems that lack the resource to 
	apply other techniques, but also that, due to its orthogonality, can be 
	combined with those in a search system.
	
	Even though it is not the main concern of this paper to address possible 
	bottlenecks that might prevent SQE to be applied in a practical context, 
	we show that it incurs in a negligible overhead. Moreover, using performance 
	techniques such as indexing the motifs, or using parallel techniques would 
	speed up the process. Performance is something that we want to look at in 
	our future research.

	In the particular case of this paper we have used Wikipedia as our test KB.  
	From the analysis of Wikipedia we have defined 2 different types of motifs 
	that we called triangular motif and square motif. To evaluate SQE we have 
	used three different datasets, Image CLEF, CHiC 2012 and CHiC 2013. The 
	first collection is used during the development of the system, while the 
	CHiC collections are used exclusively for the experiments.  The results 
	achieved by SQE are consistent for the three datasets ensuring that SQE is 
	not overfitted for a particular one. From the results, we see that the 
	triangular motif is useful to improve the results of small tops up to 
	83.87\%, a combination of the triangular and the square motifs improve the 
	result in between small and large tops up to 33.30\%, while using the square 
	motif exclusively improves the results of large tops up to 83.85\%. Also, we 
	have presented a way of combining several query graphs to improve the most 
	no matter the top to be optimized.

	In this paper we have shown the potential of using exclusively the 
	structural properties of KB to identify tightly linked entries that are 
	close semantically, with no need of using semantic analysis. We have 
	succeed in identifying the proper motifs for Wikipedia, however there are 
	many KBs and probably each has its own relevant structures. We need to 
	expand our understanding of KBs, and study what other motifs may be relevant 
	for other KBs besides Wikipedia.

	\bibliographystyle{abbrv} 
	{
		\scriptsize
		\bibliography{kdd}
	}
	\end{document}